\documentclass[12pt,preprint]{aastex}
\usepackage{emulateapj5}
\usepackage{psfig}

\shorttitle{Detection of $^{44}$Sc line emission from Cas A}
\shortauthors{J. Vink et al.}

\begin{document}

\newcommand{\subj}{Cas~A}
\newcommand{\Te}{$kT_{\rm e}$}
\newcommand{\net}{$n_{\rm e}t$}
\newcommand{\netunit}{${\rm cm}^{-3}{\rm s}$}
\newcommand{\EM}{$\frac{n_{\rm H} n_{\rm e} V}{4\pi d^2}$}
\newcommand{\NH}{$N_{\rm H}$}
\newcommand{\NHunit}{${\rm cm}^{-2}$}
\newcommand{\fluxunit}{${\rm ph/cm^2/s}$}
\newcommand{\msun}{M$_{\sun}$}
\newcommand{\spex}{{\it SPEX}}
\newcommand{\xspec}{{\it xspec}}
\newcommand{\vmekal}{{\it vmekal}}
\newcommand{\sresc}{{\it sresc}}
\newcommand{\tiff}{$^{44}$Ti}
\newcommand{\caff}{$^{44}$Ca}
\newcommand{\scff}{$^{44}$Sc}

\newcommand{\jetp}{JETP}
\newcommand{\adspr}{AdSpR}
\newcommand{\phrvl}{PhRvL}
\newcommand{\phrc}{PhRvC}

\title{Detection of the 67.9~keV and 78.4~keV lines associated with
the radio-active decay of $^{44}$Ti in Cassiopeia A}

\author{Jacco Vink\altaffilmark{1,2}, J. Martin Laming\altaffilmark{3},
Jelle S. Kaastra\altaffilmark{4}, Johan A. M. Bleeker\altaffilmark{4}, 
Hans~Bloemen\altaffilmark{4}, and Uwe Oberlack\altaffilmark{2}}

\altaffiltext{1}{Chandra fellow}
\altaffiltext{2}{Columbia Astrophysics Laboratory, Columbia University, 
MC 5247, 550 W 120th street, New York, NY 10027, USA}
\altaffiltext{3}{Naval Research Laboratory, Code 7674L, 
Washington DC 20375, USA}
\altaffiltext{4}{Space Research Organization Netherlands (SRON), 
Sorbonnelaan 2, NL-3584 CA, Utrecht, The Netherlands}

\begin{abstract}
We report the detection of the \scff\ nuclear decay lines at 67.9~keV and
78.4~keV associated with the nuclear decay of \tiff\ in Cassiopeia A. 
The line emission was observed by the PDS instrument on board BeppoSAX,
which  recently observed the supernova remnant for over 500~ks.
The detection of the line emission with a flux of
$(2.1 \pm 0.7)\ 10^{-5}$~\fluxunit\ in each line (90\% confidence)
is at the $5\sigma$ significance level, if we can assume that the 
12-300~keV continuum is adequately represented by a single power law.
However, as the nature of the continuum is not clear 
we investigate various other possibilities.
A more conservative estimate of the line flux is made by 
assuming that a power law continuum is at least a good approximation
to the continuum emission for a narrower 30-100~keV energy range.
With this limitation the measured line flux is 
$(1.9 \pm 0.9)\ 10^{-5}$~\fluxunit,
with the detection still at the $3.4\sigma$ significance level.
We suggest that together with the CGRO-COMPTEL measurement of the \caff\
line at 1157~keV 
of $(3.3 \pm 0.6)\ 10^{-5}$~\fluxunit\
a flux for all three lines
of $(2.5 \pm 1.0)\ 10^{-5}$~\fluxunit\ for Cas A can be adopted.
This implies an initial \tiff\ mass of $(0.8 - 2.5)\ 10^{-4}$~\msun.
\end{abstract}


\section{Introduction}
The detection of gamma ray line emission at 1157 keV from the supernova 
remnant Cassiopeia A by 
CGRO-COMPTEL \citep{Iyudin94} led to renewed interest in the 
nucleosynthesis and properties of \tiff, the radio-active element with which
this \caff\ nuclear de-excitation emission is associated.

However, the decay chain \tiff\ $\rightarrow$\ \scff\ $\rightarrow$ \caff\
produces two other nuclear de-excitation lines of \scff\ 
at 67.9~keV and 78.4~keV with 
a flux equal to that of the 1157~keV emission 
(see \citet{Diehl98} for a review).
As observations with the hard X-ray instruments CGRO-OSSE, RXTE-HEXTE and 
BeppoSAX-PDS failed to detect those lines \citep{The96,Rothschild97,Vink00},
some doubt was cast on the observed 1157~keV line flux, if not on the
detection itself.
With subsequent observations by COMPTEL, 
the flux estimates changed from
$(7.0 \pm 1.7)\ 10^{-5}$ \fluxunit\ 
to $(3.3 \pm 0.6)\ 10^{-5}$ \fluxunit\ \citep{Iyudin97},
whereas a 83~ks BeppoSAX-PDS spectrum narrowed the \scff\ fluxes to
a 99.7\% upper limit of $4.1\ 10^{-5}$ \fluxunit\ \citep{Vink00}.

\begin{figure*}
\centerline{
	\psfig{figure=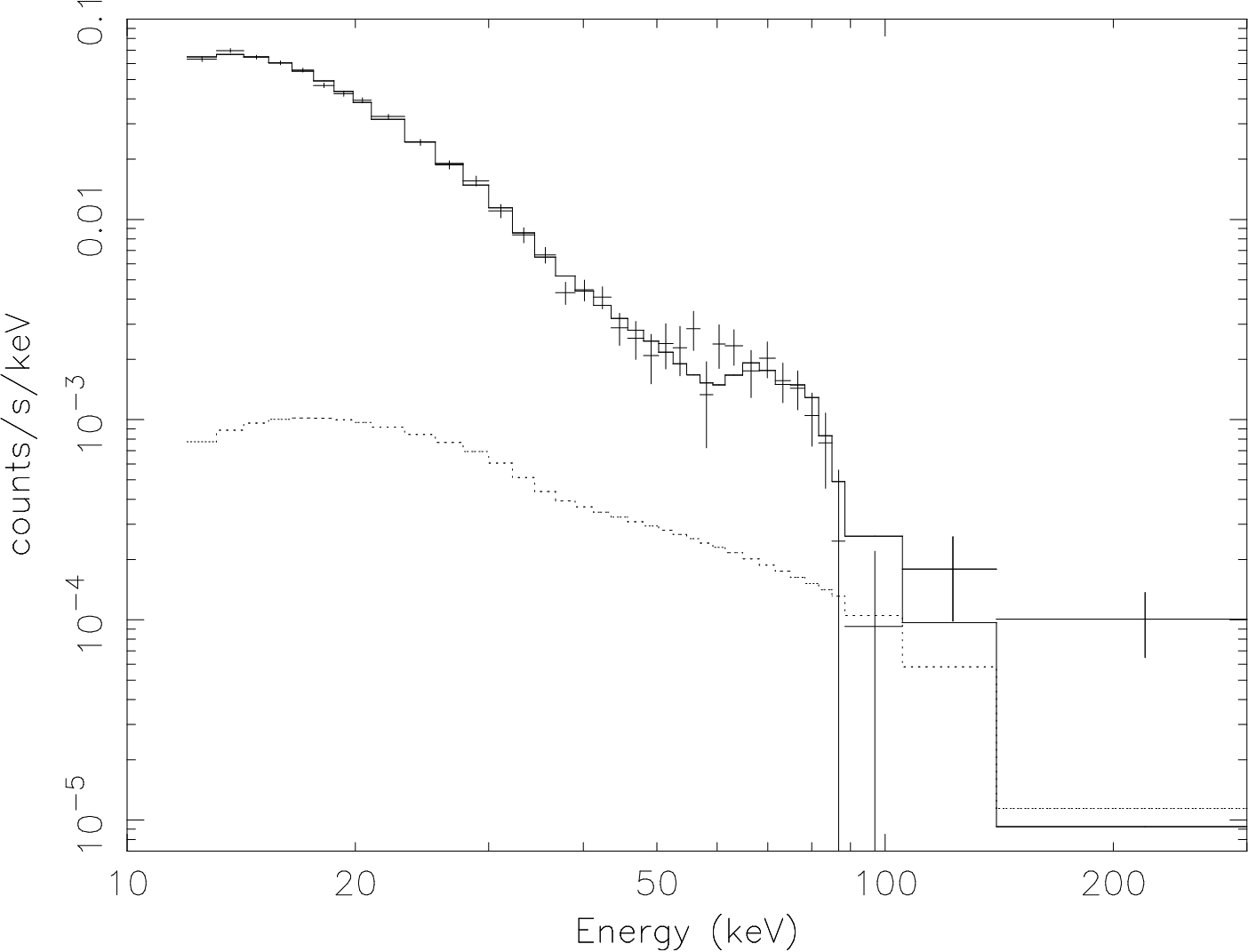,width=8.5cm,angle=-90}
	\psfig{figure=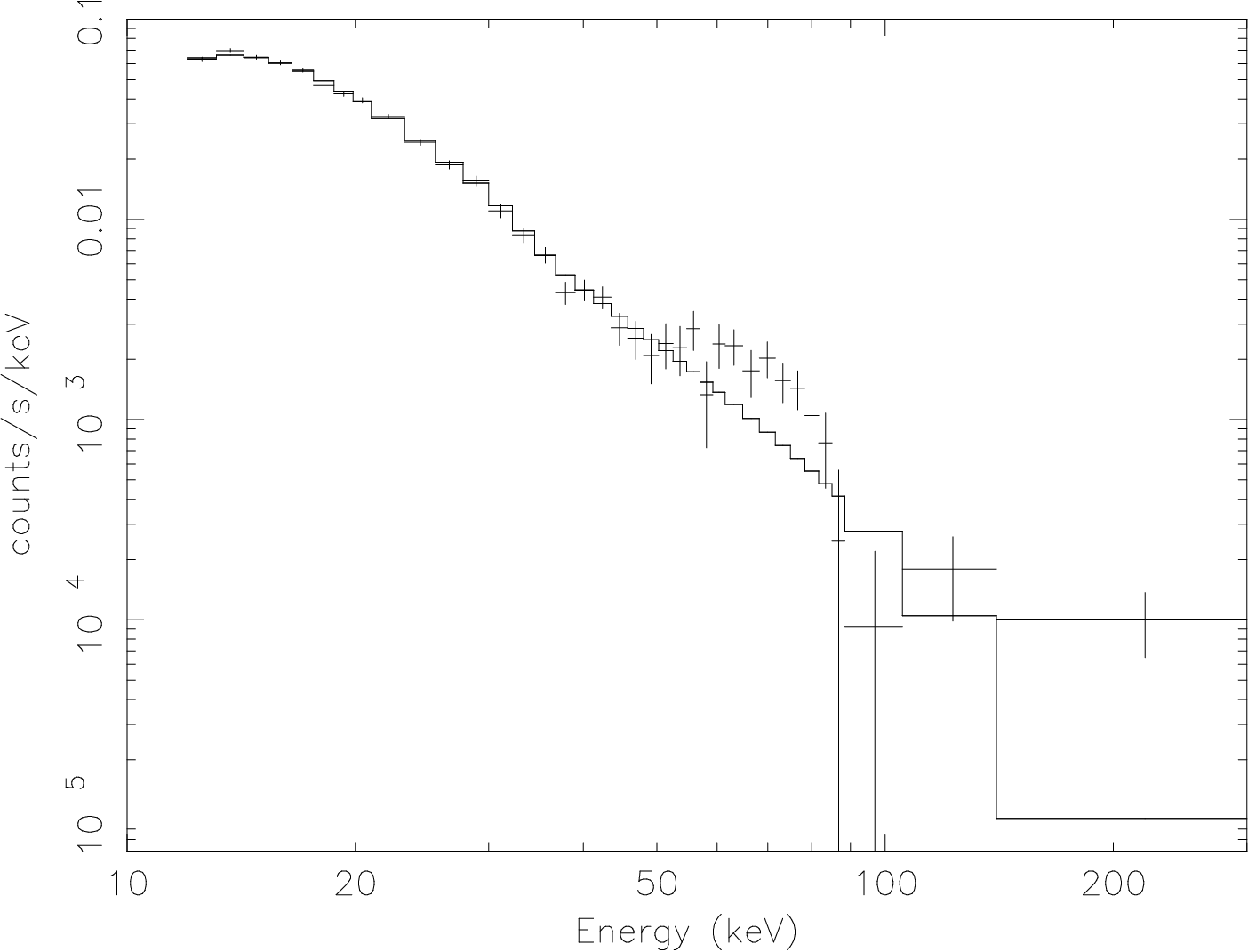,width=8.5cm,angle=-90}
}
\caption{
BeppoSAX-PDS spectrum of Cassiopeia A. The solid line in the left figure 
shows the spectral model including the \scff\ lines at 68.9~keV and 78.4~keV.
The dotted line shows the expected systematic error in the 
background subtraction, as estimated from blank field PDS spectra \citep{Guainazzi97}. 
The figure on the right shows the best fit power law model
without \scff\ lines. 
The channels above 80~keV have been rebinned for presentational purposes.
\label{fig-spec}}
\end{figure*}

In this paper we report on a new, deep, exposure of Cas A with the BeppoSAX
X-ray observatory. The BeppoSAX-PDS spectrum from this observation, together
with archival data, finally gives a reliable estimate of the
the 67.9~keV and 78.4~keV line fluxes.
\\
\\
\section{Observations and Data}
In May and June this year BeppoSAX observed Cas A for more than 500~ks. 
The primary goal
of this observation was to detect the 67.9 ~keV and 78.4~keV nuclear decay
lines with the hard X-ray Phoswich Detection System, PDS \citep{Frontera97}. 
The PDS consists of four NaI(Tl)/CsI(Na) scintillation detectors behind two 
rocking collimators with a 1.3\degr\ field of view.
During the observation the collimators switch back and forth between the on 
source position and two opposite off source positions, 
such that always one collimator observes the source and the other the 
background.
Due to the alternating position of the collimators the effective exposure
of the source with the PDS is about half the total observation time.
In this case the effective PDS exposure is 256~ks. 
We used additional archival PDS spectra of on source Cas A observations, 
resulting in a total effective PDS exposure\footnote{
The observational identification numbers for the additional archival data
used for our analysis are
30011001, 30011002 and 30795005. They were chosen on the availability of 
spectra made with the variable rise time rejection method.} of 311~ks.

As explained in the BeppoSAX analysis cookbook \citep{cookbook}, the standard
software pipeline incorporates two methods for rejection of background events
due to energetic particles.
The more advanced method, which is preferred for a background dominated
source like Cas A, uses a variable, energy-dependent,  
rise-time rejection threshold. This results in a better particle
rejection, but at the cost of a 7\% loss in effective area.
Although this is the preferred method, to get a feel of the systematic errors
involved, we also include some analysis of the same observations, but with
spectra extracted using the constant risetime rejection method. 
Compared with the statistical uncertainties,
the measured \scff\ flux does not depend
much on the method chosen, but
statistical errors
are slightly larger for spectra that were extracted using the constant risetime
method (Table~\ref{tbl-fits}).

There exists a well known error in the absolute flux density scale of
the PDS instrument of a factor $0.86\pm0.03$.
We took this factor and the 7\% effective area loss into account by multiplying
the response matrix by a factor 0.80. 
Although the statistical error of the flux calibration is small, there exists
some scatter in this relation \citep{cookbook}.
In view of this, and the small differences between measurements made
using the two rise time rejection methods,
we adopt a conservative 10\% systematic error.
In order to obtain reliable $\chi^2$ statistics, the spectra were
rebinned to channel widths of approximately $\Delta E/3$, where $\Delta E$
is the full width at half maximum (FHWM) energy resolution, which is about
9~keV at 75~keV.
Further rebinning was not necessary as the data are background
dominated, resulting in gaussian error distributions for each channel.
We verified that the spectra of each collimator were consistent with
the best fit model to the overall spectrum (see below).
Similarly, we verified the consistency of the recently observed spectra and
the archival spectra.

As far as we know there are no instrumental background lines
coinciding in energy with the \scff\ lines. Possible 
contamination sources of line emission close to the \scff\ lines
are tantalum from the collimator
and the on board $^{241}$Am calibration source.
$^{241}$Am emits photons at 60~keV, whereas the Ta K$\alpha$ and
K$\beta$ transitions are at 57~keV and 65~keV with an emission ratio of 4:1.
These lines
may be responsible for a small emission excess around 60~keV (Fig.~\ref{fig-spec}).
However, adding these lines to our models changed the measured \scff\ line flux
with less than 5\%; much smaller than the statistical uncertainty in the
line flux.

Note that the observed power law normalization of the PDS 
spectrum agrees with that of the RXTE-HEXTE \citep{Allen97}.
At 15.9~keV their model fits imply $2.6\, 10^{-4}$~ph/s/keV, 
whereas the PDS spectrum indicates $2.5\, 10^{-4}$~ph/s/keV. 
The photon index above
16~keV of the RXTE-HEXTE spectrum, $3.04\pm0.15$, 
agrees within $1.6\sigma$\ with the value reported below.
The continuum normalization also agrees within $1.5\sigma$ with 
measurements by CGRO-OSSE \citep{The96},
which measured a flux density of 
$(9.0\pm 2.1)\ 10^{-7}$~ph/cm$^{2}$/keV at 100~keV, compared
to $(5.8\pm 0.8)\ 10^{-7}$~ph/cm$^{2}$/keV for our measurement,
obtained from extrapolating the power law normalization to
100~keV (first row in Table~\ref{tbl-fits}).
Note that the CGRO-OSSE had a much larger field of view 
($11\degr\times4\degr4$) than the BeppoSAX-PDS.

\section{Data analysis}
We present the combined PDS-spectrum and the simplest possible model, 
consisting of a power law continuum and two nuclear decay lines, 
in Figure~\ref{fig-spec}.
For this modeling of the continuum shape the \scff\ lines are detected with a
significance of more than $5\sigma$ ($\Delta \chi^2 = 30$).
The flux in each line is $(2.1 \pm 0.7)\ 10^{-5}$ \fluxunit\ and the photon 
index is 3.3 (see Table~\ref{tbl-fits} for details).
However, the nature of the hard X-ray continuum is still under debate.
It has been argued that it is synchrotron radiation from shock accelerated
electrons, or, alternatively, 
that it is bremsstrahlung emission caused by a non-thermal tail to the 
thermal electron distribution 
\citep{Asvarov90,Allen97,Favata97,Bleeker01}.
A more specific bremsstrahlung model was worked out by 
\citet{Laming01a,Laming01b},
who calculated the spectrum of electrons accelerated by lower hybrid waves 
associated with shocks in the ejecta.

In this letter we limit ourselves to the detection of the
\scff\ lines. We therefore leave a detailed discussion of the hard X-ray 
continuum, which should also include data for the energy range 0.5-10~keV, 
to a future article. However, in order to test the dependence of the measured 
line flux estimates on the assumed continuum shape we tested various continuum
models, the results of which can be found in Table~\ref{tbl-fits}.
It is surprising that the best fit to the PDS-spectrum in a statistical sense
is provided by a power law model plus \scff\ line emission.
However, as the measured line flux depends on the continuum model chosen,
we have to take the uncertainty about the nature of the continuum into 
account. 

\begin{figure*}
\hbox{
\parbox{9.2cm}{
\psfig{figure=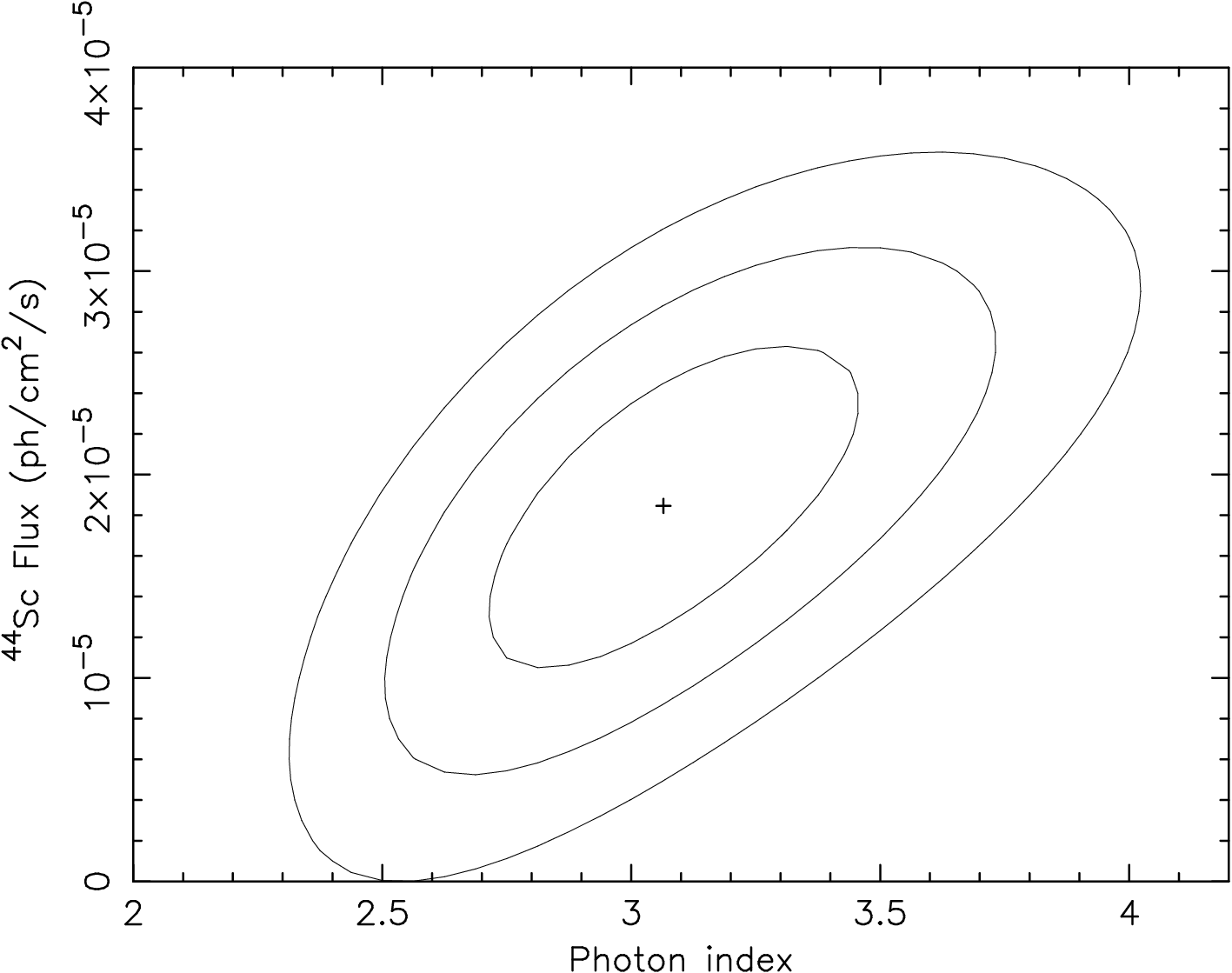,width=8.9cm,angle=-90}}
\hskip 0.5cm
\parbox{7cm}{
\vskip 4.5cm
\caption{
Confidence ellipses for the combination of power law index and \scff\ flux for the spectral energy range of 30 keV to 100~keV. The contours are 1, 2 and 3$\sigma$ 2-parameter confidence levels ($\Delta \chi^2 = 2.3, 6.17, 11.8$, see \citet{Lampton76}). \label{fig-contour}}}}
\end{figure*}

It is clear from spectral fits to the narrower 30-100~keV spectral 
range that the inclusion of a thermal component with 
\Te = 4.2~keV has little
effect on the estimated \scff\ flux.
Certainly, for this narrow range a power law is a reasonable approximation
for the continuum. As the energy range is smaller, there is more statistical
uncertainty about the photon index, 
and therefore the \scff\ line flux is more uncertain. 
Indeed the significance of the line emission drops from the $5\sigma$ to
the $3.4\sigma$ level.
The photon index versus line flux confidence contours for this 
energy range is shown in Figure~\ref{fig-contour}.
The $3\sigma$ upper limit on the flux in both lines is 
$3.5\ 10^{-5}$ \fluxunit, based on the 30-100~keV energy range.
This upper limit is comparable to the \caff\ line flux at 1157~keV 
recently obtained from CGRO-COMPTEL data, 
$(3.3 \pm 0.6)\ 10^{-5}$~\fluxunit\ (68\% confidence range),
see \citet{Iyudin97}.
However, both measurements are consistent given the 
25\% systematic uncertainties for the COMPTEL measurement \citep{Dupraz97}.

Although it is difficult to combine the COMPTEL and the PDS results, due to 
the unknown nature of the systematic errors, the fact that there are
now two independent measurements of line emission associated with the
\tiff\ decay increases the credibility of the detections.
We therefore suggest adopting a \scff/\caff\ line flux for Cas A of 
$(2.5 \pm 1.0)\, 10^{-5}$~\fluxunit, 
which is consistent with both the PDS and COMPTEL measurements.

\section{Conclusions}
Since the initial discovery of the \caff\ nuclear decay lines by 
CGRO-COMPTEL \citep{Iyudin94}, our knowledge of the formation and decay 
of \tiff\ has substantially improved.
Recently, the decay time of \tiff\ has been accurately measured to be 
$85.4 \pm 0.9$~yr by three independent experiments 
\citep{Ahmad98,Goerres98,Norman98}.
As the Cas A supernova was probably observed by the English astronomer
J. Flamsteed in A.D. 1680 \citep{Ashworth80}, 
the age of Cas A is $\sim$ 320~yr. 
A good alternative is to use the kinematic age of the fast moving optical 
knots, 330~yr \citep{Thorstensen01}, but the age difference is so small
that it has little effect on the inferred range of initial \tiff\ mass.
The line flux of $(2.5 \pm 1.0)\, 10^{-5}$~\fluxunit, 
combined with the \tiff\ decay time and
a distance to Cas A of $3.4^{+0.3}_{-0.1}$~kpc \citep{Reed95},
yields an initial \tiff\ mass in the range $(0.8 - 2.5)\ 10^{-4}$~\msun, with
$1.2\ 10^{-4}$~\msun\ corresponding to the adopted \scff/\caff\ flux and
distance to Cas A.

This is rather high compared to model predictions, which usually indicate
\tiff\ masses below $10^{-4}$~\msun,
except for progenitor masses around 12\msun\
and above $\sim$25\msun\ \citep{Timmes96}. 
For that reason \citet{Mochizuki99} made the interesting suggestion that 
the \tiff\ decay may be delayed, as a result of complete ionization of
\tiff,
inhibiting the decay of \tiff\ by the capture of a K-shell electron.
However, recently \citet{Laming01c} showed that the electron temperature in
the ejecta was probably never high enough to seriously affect the \tiff\ decay.

The production of \tiff\ increases with the size of the helium core of the 
progenitor, but material falling back on the neutron star or
black hole limits the amount of ejected \tiff.
Model calculations show that massive stars have more fall back, 
but pre-supernova wind loss may limit the amount of material falling back
\citep{Timmes96}. 
This agrees with the idea that the progenitor of Cas A was
a not too massive ($\sim$ 30\msun) Wolf-Rayet star that suffered heavy mass
loss \citep{Vink96}. Additionally, assymetries in the explosion may
have increased the amount of \tiff\ synthesized \citep{Nagataki98}.
The likely presence of a neutron star \citep{Chakrabarty01} in Cas A further 
constrains the explosion scenario for Cas A, as too much fall back would have
resulted in the formation of a black hole. 

Although we have now finally detected the \scff\ nuclear decay lines
at 67.9~keV and 78.4~keV, interesting observations remain to be done
with future hard X-ray and Gamma-ray  missions. 
The solid state detectors
on board Integral will be able to measure the line broadening of the \caff\
line accurately, and constrain the properties of the hard X-ray continuum
further.
In addition, hard X-ray experiments using multi-layer mirrors
will be able to map the spatial distribution of \tiff\ in Cas A on the
arcminute scale.

\acknowledgments
Support for this work was provided by the NASA
through Chandra Postdoctoral Fellowship Award Number PF0-10011
issued by the Chandra X-ray Observatory Center, which is operated by the
Smithonian Astrophysical Observatory for and on behalf of NASA under contract
NAS8-39073.
JML was supported by basic research funds of the Office of Naval Research
This research has made use of SAXDAS linearized and cleaned event  
    files (Rev.2.1.4) produced at the BeppoSAX Science Data Center.

\begin{deluxetable}{lccccccccc}
\tabletypesize{\scriptsize}
\tablecaption{Summary of spectral model fits. 
\label{tbl-fits}}
\tablewidth{0pt}
\tablehead{
RT & model  & spectral & \scff\ flux         & 
PL  &  PL norm &  radio & roll off & Emission 
	& $\chi^2/\nu$ \\
method &  & range & & index & @ 1 keV & index & energy & measure\tablenotemark{a} \\
                  &          & (keV)  & ($10^{-5}$~\fluxunit) & 
& (ph/s/keV) & & (keV) & ($10^{12}$~cm$^{-5}$) \\
 & & & & & & &
}
\startdata
V & PL  &12 - 300 & $2.1 \pm 0.7$ & $3.30 \pm 0.05$ & $2.3 \pm 0.3$ && && 
	69.5/63\\

V & PL+therm.& 
	12 - 300 & $1.0 \pm 0.7$ & $2.71\pm0.06$ & $0.28 \pm 0.05$ & 
	& & 9.25\tablenotemark{b} (fixed) &  96.7/63\\

V & PL+therm.& 
	12 - 300 & $2.0 \pm 0.7$ & $3.2 \pm 0.2$ & $1.8\pm0.8$ & & & $1.3 (< 4.5)$ & 69.4/62\\

V & \sresc+therm. & 
	12 - 300 & $3.2 \pm 0.8$ & & & 0.85 ($> 0.848$)\tablenotemark{c} & $9.2 \pm 0.3$ &  9.25\tablenotemark{b} (fixed) & 120.8/63\\

V & \sresc+therm. & 
	12 - 300 & $2.8 \pm 0.9$ & & & 0.85 ($> 0.838$)\tablenotemark{c} & $10 \pm 1$ & $5.8 \pm 0.9$  & 75.9/62\\

\\
V & PL &30 - 100 & $1.9\pm 0.9$  & $3.1 \pm 0.4$ & $0.9^{+3.6}_{-0.7}$ & & &&
	15.6/20\\
V & PL+therm.& 
	30 - 100 & $1.8 \pm 0.9$ & $3.0 \pm 0.04$ & $0.7^{+2.8}_{-0.6}$ & & & 
	9.25 (fixed) & 15.5/20\\

\\
C & PL & 12 - 300 & $2.5 \pm 0.8$ & $3.33 \pm 0.06$ & $2.5 \pm 0.4$ && & & 74.5/63\\
C & PL &30 - 100 & $1.7\pm 1.0$  & $2.8 \pm 0.5$ & $0.30^{+1.51}_{-0.24}$ & 
	& & & 18.7/20\\
\enddata
\tablecomments{
All spectral models can be found in X-ray spectral analysis package 
\xspec\ v11 \citep{xspec}. 
The first column indicates which background rejection criterium was used, 
- variable (V), or constant (C) - to obtain the source spectrum (see text).
The \scff\ lines were modelled by two deltalines with energies fixed
at 67.9~keV and 78.4~keV, and equal line flux.
PL is short for power law. We used \vmekal\ (a collisional
equilibrium model) for the thermal component \citep{mekal}. 
The \sresc\ model is 
described in \citet{Reynolds99}; it gives the synchrotron emission from
a relativistic power law distribution of particles with an exponential 
cut-off. 
Errors are statistical errors, and correspond to 90\% confidence intervals.
}
\tablenotetext{a}{The emission measure is defined as \EM. Note that
in order to keep in line with models used by \citet{Vink96}, \citet{Favata97} 
and \citet{Vink00} we used a plasma enhanced in helium and nitrogen 
by a factor 10 (based on optical observations of the shocked circumstellar
medium, e.g. \citet{Chevalier79}).
The electron temperature was fixed to \Te = 4.2~keV \citep{Vink00}.
Note that the ejecta is most likely the main source of thermal X-ray
emission in Cas A, and so an O-rich composition could also be a 
reasonable model. This would not, however, make any difference to the
shape of the thermal continuum at photon energies relevant here.
}
\tablenotetext{b}{The fixed emission measure is identical to that used 
by \citet{Vink00}, comparing with ASCA-SIS0 and BeppoSAX-MECS data 
shows that this implies that below 10~keV about half of the continuum is 
thermal, whereas the other half comes from an additional component.}
\tablenotetext{c}{
We allowed the radio spectral index, nominally 0.77 for Cas A, 
to vary in the range 0.7 to 0.85;
the normalization parameter is the radio flux density at 1~GHz, 
which was fixed to 2720~Jy \citep{Green-cat}. 
The parameters derived from this model should be treated with caution, as
the values for radio index and roll off energy are highly correlated.
}

\end{deluxetable}

\end{document}